\documentclass{article}

\usepackage{amsmath}
\usepackage{PRIMEarxiv}

\usepackage[utf8]{inputenc} % allow utf-8 input
\usepackage[T1]{fontenc}    % use 8-bit T1 fonts
\usepackage{hyperref}       % hyperlinks
\usepackage{url}            % simple URL typesetting
\usepackage{booktabs}       % professional-quality tables
\usepackage{amsfonts}       % blackboard math symbols
\usepackage{nicefrac}       % compact symbols for 1/2, etc.
\usepackage{microtype}      % microtypography
\usepackage{lipsum}
\usepackage{fancyhdr}       % header
\usepackage{graphicx}       % graphics
\graphicspath{{media/}}     % organize your images and other figures under media/ folder

\title{Logistic regression with missing responses and predictors: a review of existing approaches and a case study
}
\author{
Susana Rafaela Martins* \\
 Escola Superior de Desporto e Lazer\\
      Instituto Politécnico de Viana do Castelo\\
      Portugal\\
      \texttt{srgm@estg.ipvc.pt}
    \And
 Jacobo de Uña-Álvarez, María del Carmen Iglesias-Pérez\\ 
Department of Statistics and OR\\
Universidade de Vigo\\
Spain\\ \texttt{jacobo@uvigo.es, mcigles@uvigo.es}
}

\begin{document}
\maketitle

\maketitle

\begin{abstract}
In this work logistic regression when both the response and the predictor variables may be missing is considered. Several existing approaches are reviewed, including complete case analysis, inverse probability weighting, multiple imputation and maximum likelihood. The methods are compared in a simulation study, which serves to evaluate the bias, the variance and the mean squared error of the estimators for the regression coefficients. In the simulations, the maximum likelihood and multiple imputation methodologies present the smallest mean squared errors, their relative performance depending on the particular missing mechanism. The methods are applied to a case study on the obesity for schoolchildren in the municipality of Viana do Castelo, North Portugal, where a logistic regression model is used to predict the International Obesity Task Force (IOTF) indicator from physical examinations and the past values of the obesity status. All the variables in the case study are potentially missing, with gender as the only exception. The results provided by the logistic regression indicate the relevance of the past values of IOTF and physical scores for the prediction of obesity. Practical recommendations for those handling missing data are given. 
\end{abstract}

\begin{keywords}
Inverse probability weighting; maximum-likelihood; multiple imputation; physical performance; obesity
\end{keywords}

\section{Introduction}
Overweight problems and a sedentary lifestyle, particularly in children, are nowadays a large concern in society. Industrialized and processed food, as well as lack of exercise, have been two of the biggest causes of health problems, especially in child obesity.
The obesity is a global scourge because worldwide obesity has tripled since 1975. In most countries, obesity and being overweight kill more people than being underweight. According to the World Health Organization (WHO), in 2019 over 340 million children and adolescents aged 5-19 were overweight or obese \cite{WHO21}.

The Body Mass Index (BMI) is the standard measure to assess adult obesity levels. The BMI, formerly called the Quetelet index, is a measure for indicating nutritional status in adults. WHO defines the BMI as the person’s weight in kilograms divided by the square of the person’s height in metres \cite{WHO20}.  In children and adolescents however, this is not so simple. The author \cite{Jane} argues that the age, growth rate and puberty have a considerable influence on the amount of fat that is required to be available at any given time. Consequently, creating an overarching standard that define overweight or obese for all ages is difficult. This has motivated the appearance of different reference indices. According to \cite{Valerio}, the  International  Obesity  Task  Force (IOTF) system is more biologically meaningful compared to the references based on percentiles. The IOTF system uses smooth sex-specific BMI curves constructed to mach the values of overweight and obesity at 18 years, thus identifying the age and gender BMI cut-offs from overweight and obesity. These cut-off points are based on large data sets covering different races and ethnicities.  Like other countries of Europe, Portugal employs two different international systems to classify overweight and obesity, IOTF being one of these \cite{Hlopes}.

The study \cite{LP2006} performed a morphofunctional analysis of the schoolchildren from Viana do Castelo in which the IOTF index was registered. In this paper we consider information from that study, corresponding to $n=230$ children. The collected data refer essentially to three types of individuals’ characteristics: social and demographic characteristics, morphological characteristics and physical fitness characteristics. These variables were collected according to the AAHPERD protocols \cite{AAPHER76,AAPHER80} and  the EUROFIT tests \cite{EF1988}, annually between 1997 and 2000, and then six years later, in 2006. The social and demographic variables were recorded only at the first moment, that is, in the year the individual entered the database. The physical and morphological variables were recorded longitudinally. Our goal is to study the relationship between the IOTF and the physical performance of children, so an early screening of obesity becomes feasible. Ideally, such a screening should result from a medical follow-up. However, children do not visit the medical doctor as often as they attend sports classes, so the relationship between IOTF and physical variables becomes important.

One issue in the aforementioned case study is that of missing data. Missing data is a frequent problem in statistical studies, particularly those of a longitudinal nature, where it may be difficult or even impossible to register the target variables for each individual along time. Methods for handling missing data have received a lot of attention in the statistical literature; see for instance \cite{Ibrahim2005, ibrahim2012missing,ibrahim2009missing, enders2013dealing,enders2017multiple, carpenter2021missing, Wang}. Specifically, Ibrahim \cite{Ibrahim2005, ibrahim2012missing, ibrahim2009missing} considered the following methods for inference in generalized linear models with missing covariates: complete case analysis (CC), maximum likelihood (ML), multiple imputation (MI), fully Bayesian (FB) and inverse probability weighting (IPW). The CC analysis has the advantage of being simple, but it may be inefficient since it discards the individuals with missing records. The IPW method proceeds by estimating the probability of non-missingness, so its inverse can be used to weight the data. While IPW performs well for a single predictor, is less efficient in the case of multiple missing variables \cite{Horton}. MI is a simulation-based methodology, which recreates the missing values as similar as possible to the (unavailable) real values. See \cite{Horton,Ibrahim2005,ibrahim2009missing, ibrahim2012missing,Bori,erler2016,erler2019} for more details.

Carpenter et al. \cite{carpenter2006comparison} conclude that MI and IPW provide unbiased estimates under relatively weak assumptions in linear regression. These authors indicate that the performance of both methods is similar, although the MI methodology is slightly more efficient than the IPW. More recently, \cite{carpenter2021missing} investigated the relative performance of MI, IPW and ML procedures, with a focus on sensitivity analysis. One conclusion is that ML and MI generally outperform IPW. Additionally, Allison \cite{allison2012handling} argues that ML is, in general, more efficient than MI. Besides, the result of ML is always the same while as mentioned, with MI there exists an extra variability due to the randomness inherent of the method. It is also noteworthy that MI may present a potential conflict between the imputation model and the model under analysis, which is not the case with ML \cite{allison2012handling}. 

In this work a logistic regression is performed to model the relationship between the IOTF and other variables in the Viana do Castelo study. Specifically, the aim is to predict the IOTF in 2006 given the IOTF and physical performance in 2000, while controlling for gender. The IOTF was considered as a dichotomous variable, indicating overweight (1) and normal weight (0). Physical performance was represented by the 60-second sit-ups (ABD), which evaluates the number of sit-ups that the child is able to do in 60 seconds. Both the response variable and covariates are subject to missingness, with the only exception of gender. Therefore, several of the aforementioned methods are first investigated in simulated scenarios which mimic the Viana do Castelo case study. Our research relates \cite{Ibrahim2005,peng2008,Raghunathan,carpenter2021missing} in that logistic regression with missing data is considered; however, in these papers missingness affects the covariates or the response variable, while in our setting missing data in both the covariates and the response may occur. Therefore, our simulation study brings new insights on the relative performance of CC, IPW, MI and ML methods in complex scenarios of missing data.

The rest of the paper is structured as follows. Section 2 gives a description of different methods to deal with missing data: CC, IPW, MI and ML. In Section 3 a simulation study is conducted, including four different scenarios with missing observations in both the response variable and the covariates. Section 4 provides the application of the aforementioned methods to the Viana do Castelo study. Finally, Section 5 gives the conclusions of our research and some practical recommendations.

\section{Methods}

\subsection{Logistic Regression}
The  logistic regression model is an adaptation of the linear regression model for binary responses. In the Viana do Castelo study, the logistic regression model is used to relate the IOTF to a vector of covariates. Similarly as in \cite{weiss}, let $Y_{i}$ be the response variable, $Z_{i}$ a $q\times1$ vector of covariates, and $\beta$ the $q\times1$ unknown parameter vector,$i\in\{1, .., n\}$. 
The logistic regression model states that, conditionally on $Z_i$,
\begin{equation*}
 Y_{i}\sim Bernoulli (\mu_{i})
\end{equation*}
where $\mu_{i}=P(Y_{i}=1|Z_{i})$ depends on the linear predictor $\eta_{i}=\beta^{T}Z_{i}$ through the logit link function. Specifically,
\begin{equation*}
\mu_{i}=\frac{\exp(\beta^{T}Z_{i})}{1+\exp(\beta^{T}Z_{i})}.
\end{equation*}

For complete data, the estimation of the $\beta$ parameter vector is based on the maximum likelihood principle. However, estimation is not so immediate in the presence of missing data.

\subsection{Mechanisms of Missing Data}

Similarly to \cite{Molenberghs} and \cite{Chen}, let $(X_{i}, X_{i}^{mis})$ be formed by the elements of $(Y_{i}, Z_{i})$ such that $X_{i}$ is a $d$-dimensional non null vector with $0<d\leq q$ that is observed for all $i$'s, while $X_{i}^{mis}$ is the vector with the observations that may or may not be available for some $i$'s. Let $R_{i}$ be the indicator of missing values, that is,
 \begin{equation}
   R_{i}=\left\{\begin{array}{ll} 1,\ \ if \ \ X_{i}^{mis}\ \ is \ \ observed \ \ 
   \\0,\ \ if\ \ X_{i}^{mis} \ \ is \ \ missing  \end{array} \right.
 \label{ri}
\end{equation}

The data are missing completely at random (MCAR) when the probability of missing values is unrelated to either the specific values. Thus, under MCAR, the missingness probability is given by
 \begin{equation*}	
P(R_{i}=0|Y_{i},Z_{i})=P(R_{i}=0|X_{i},X_{i}^{mis})=P(R_{i}=0).
\end{equation*}

The data are missing at random (MAR) when the probability of missing values is related to the set of observable variables. Under MAR, the probability of missingness is
 \begin{equation}	
P(R_{i}=0|Y_{i},Z_{i})=P(R_{i}=0|X_{i})=P(X_{i}).
\label{eqMAR}
\end{equation}

 The other mechanism of missing values is not missing at random (NMAR). Under NMAR the probability of missingness depends on outcomes which are not always available. In other words, this probability is related to $X_{i}^{mis}$ because it depends on at least some components of $X_{i}^{mis}$:
\begin{equation*}	
P(R_{i}=0|Y_{i},Z_{i})=P(R_{i}=0|X_{i},X_{i}^{mis}).
\end{equation*}

\subsection{Methods for missing data}

There are several methods to analyze missing data. In this article we will present complete case analysis (CC), inverse probability weighting (IPW), multiple imputation (MI) and maximum likelihood (ML).

\subsubsection{Complete Case Analysis}
 
 Complete case analysis is a widely used technique for handling missing outcomes. In this methodology, individuals who have missing values in any of the variables under analysis are discarded. This method is simple but generally inefficient, since it may lead a substantial reduction of the sample size. The fact that individuals are removed when a single variable is missing makes this worse. Additionally, when the missing mechanism depends on the outcome variable $Y$, complete case analysis may lead to biased estimators \cite{carpenter2021missing}.

 Basically there are two situations in which the estimators given by CC are consistent. One of the situations is when we are faced with an MCAR scenario, that is, when the probability of missing data is independent of the covariates and the response variable. The other situation occurs when we are faced with a MAR scenario, where the probability of missing data is independent of the response variable and the regression model is well defined. Except in these two situations, CC estimators are generally inconsistent \cite{Seaman}.

\subsubsection{Inverse probability weighting}
The IPW is one of the methodologies used to correct the bias produced by CC. Similarly to CC, IPW only uses only complete data. However, each observation is given a different weight in order to overcome the observation bias. The weights of the observations are generally unknown, so it is necessary to estimate them. Generally, a logistic regression model is fitted whose response variable is R (missing indicator), called the missing model \cite{Seaman}.

%On \cite{carpenter2006comparison} IPW is also defined as a method whose main objective is to eliminate bias, resorting to the reconstruction of the population by weighting the data of individuals with less probability of being observed. We review this idea in the scope of logistic regression with missing observations. Let $(Y_{i}, Z_{i})$ a random vector of individuals, $i\in\{1, .., n\}$, where $Y_{i}$ is the response variable and $Z_{i}$  the vector of covariates. 

For complete data, the maximum likelihood estimator of $\beta$ for the logistic regression model is the solution of the equation
\begin{equation}
   \sum_{i=1}^{n} {Z_{i}}(Y_{i}-\mu_{i})=0
   \label{ipw}
\end{equation}	
where $\mu_i=\exp(\beta^t Z_{i})/(1+\exp(\beta^t Z_{i}))$ and $\beta$ is the column vector of regressions coefficients. At the true value of the regression coefficient the expectation of the left-hand side in (\ref{ipw}) is 0, and this ensures that the parameter estimates are consistent.
Let $R_{i}$ the indicator of missing values as introduced in equation (\ref{ri}). The observed data estimating equations can be written as
  
\begin{equation*}
   \sum_{i=1}^{n} R_{i}Z_{i}(Y_{i}-\mu_{i})=0.
\end{equation*}	

If weights are used, based on estimated non-missingness probability $\pi_i=1-P(X_i)$ with $P(X_i)$ as defined in (\ref{eqMAR}) for MAR scenario, one has
\begin{equation}
   \sum_{i=1}^{n} \frac{R_{i}}{\pi_{i}} Z_{i}(Y_{i}-\mu_{i})=0.
   \label{ipw_peso}
\end{equation}

As before, the left-hand side of (\ref{ipw_peso}) is zero-mean; thus, the solution to the score equation (\ref{ipw_peso}) is consistent for $\beta$. 

The IPW estimators are consistent, which represents an advantage of this methodology over the CC. A disadvantage of this methodology is the fact that there is no guarantee of bias correction if the weighting model is incorrectly specified. In this case the estimate may be inconsistent \cite{Seaman}.

\subsubsection{Multiple imputation}

Multiple imputation is a methodology to address missing data. According to \cite{MolenberghsHandbook}, the purpose of multiple imputation is to generate values for the missing outcomes, so multiple complete data sets are obtained.
%This technique is based on simulations and provides a measure of uncertainty about what missing data is predicted based on the observed data 
This technique is based on simulations where missing data is predicted based on information from observed data \cite{peng2008}.

Imputed data must preserve the integrity, uncertainty, and knowledge about the generation of the original data. For this, there are two multivariate approaches: Joint Modeling (JM) and Fully Conditional Specification (FCS). JM based on parametric statistics uses imputation procedures with well-defined statistical properties. On the other hand, FCS provides a semiparametric and flexible alternative, establishing a multivariate model through a series of conditional models \cite{Buuren}.

According \cite{Horton,MolenberghsHandbook,peng2008} MI is essentially based on three stages: imputation, analysis and grouping.
In the first step, several possible values are assigned to the missing observations. These values are imputed according to the uncertainty associated with the model underlying the missing data. The number of imputations does not have to be very high. In general, sets with 5 or 10 imputations give very satisfactory results.
In the second stage, each of these data sets, obtained by imputation, are analyzed using typical methodologies for complete data. In the last step, the results are pooled into a single set of parameter estimates and standard errors.

In general, the focus is the inference of a parameter $\beta$ that characterizes a parametric model, generically denoted as $P(V|\beta)$, where $V=(Y, Z)=(X, X^{mis})$. In the setting of multiple imputation, the analyses for parameter $\beta$ requires the calculation of the posterior distribution $P(\beta|X,R)$.
In the MAR scenario, using conditional probability the posterior distributions could be evaluated at each of the imputed values of $X^{mis}$, as:
\begin{equation*}
   P(\beta|X)\approx \frac{1}{m} \sum_{k=1}^{m} P(\beta|X, X_{k}^{mis}),
   \label{discreto model miss}
\end{equation*} where $m$ is the number of imputations \cite{MolenberghsHandbook, carpenter2006comparison}.

In practice, for multiple imputation to be valid, we must include in the imputation model all the variables that are needed to ensure that the response is missing at random. When relevant variables are excluded, a bias in the estimates may occur. A practical advantage of MI is that the computational algorithm of imputation is remarkably stable, and apparently reasonable results can generally be obtained even when the number of variables is high \cite{carpenter2006comparison}.  Generally, MI is more frequently used than IPW; when the imputation model is correct, the estimates provided by MI are consistent. Alternatively, IPW is attractive because it is based on a simple idea for correcting bias, which is easily implemented with any conventional software. However, one of the difficulties of these methods is the fact that it is necessary to estimate the non-missing probability model.

In summary, it can be said that MI is a simulation-based methodology, whose objective is to deal with missing data in order to obtain a valid statistical conclusion. This methodology has the same optimal properties as ML (see below), removing some of its limitations, and can be used with any type of data. In the case of MAR data, MI can obtain consistent, asymptotically efficient, and asymptotically normal estimates \cite{Bori, allison2001}.

\subsubsection{Maximum Likelihood}

Maximum likelihood is another approach for dealing with missing data that assumes a MAR scenario. Based on the same assumptions as MI, both ML and MI produce consistent, asymptotically efficient and asymptotically normal estimates \cite{allison2001}.

ML focuses on the parameters $\beta$ of the conditional distribution regression $f(Y|Z,\beta)$. Regardless of the presence or absence of missing data, the first step in ML is the construction of the likelihood function. When some variables are missing, the information can be recovered by estimating the distribution of covariates\cite{Ibrahim2005}.  This distribution is maximized through an expectation-maximization algorithm. For each missing observation, multiple entries are created with all possible values. The probability of observing these hypothetical results is calculated based on the observed data. This augmented full data set can be used to fit the regression model.
When there are many variables, some simplification of the joint distribution is often needed. Ibrahim and Horton \cite{Ibrahim2005, Horton} suggest a conditional approach. For example, consider the case of three variables $Y$, $Z_{1}$ and $Z_{2}$. A factored regression model consist of a sequence of univariate conditional models $P(Y|Z_{1},Z_{2})$, $P(Z_{1}|Z_{2})$ and $P(Z_{2})$ such that the joint distribution can be factorized as 
\begin{equation*}
   P(Y,Z_{1},Z_{2}) = P(Y|Z_{1},Z_{2}) P(Z_{1}|Z_{2}) P(Z_{2}).
   \label{prob ML}
\end{equation*} 
Note that all of the variables can contain missing values. These missing values are integrated out by posing a distribution assumption for each variable subject to missingness \cite{Horton}.
%Complications can arise in the application of the algorithm of maximum expectation, as well as in the calculation of the standard errors of the estimates .

\section{Simulation study}

A simulation study was carried out to understand which method is the most appropriate to estimate the model parameters that explain the variable of interest in the Viana do Castelo study. For this, the simulation design imitated our real data application. We simulated 1000 data sets with three different dimensions $n=\{230,400,1000\}$ by considering four variables: the variable of interest $Y$ and three covariables $Z_{1}$, $Z_{2}$ and $Z$. Firstly, we generated the full data. Secondly, we generated missing values according to four scenarios. Thirdly, estimates for the regression coefficient were obtained and the results were analysed.

\subsection{Simulation design}

\vspace{0.5cm}
\textbf{Full data}
\vspace{0.25cm}

To simulate the full data, the variable $Z$ was drawn from a Bernoulli distribution $Ber(p_1)$ with $p_{1}=0.55$. Conditioned to $Z=z$,  $Z_{1}$ was drawn from a $Ber(p_2(z))$ with $p_{2} (z)$,  where $p_{2}(0)=0.27$  and $p_{2}(1)=0.31$. The variable $Z_{2}$ was drawn conditioned to $Z=z$ and $Z_{1}$ from $N(\mu, \sigma)$, where:
         \\ for $z = 0$, $Z_{1} = 0$, we took $\mu=31$; $\sigma=7$;
         \\ for $z = 0$, $Z_{1} = 1$, we took $\mu=25$; $ \sigma=10$; 
         \\ for $z = 1$, $Z_{1} = 0$, we took $\mu=33$; $ \sigma=6$;
         \\ for $z = 1$, $Z_{1} = 1$, we took $\mu=28$; $ \sigma=9$.

The $Y$ was drawn from $Ber(p_{3})$ where
    \begin{equation}
        p_{3}=\frac{exp(b_{0}+b_{z}Z+b_{1}Z_{1}+b_{2}Z_{2})}{1+exp(b_{0}+b_{z}Z+b_{1}Z_{1}+b_{2}Z_{2})}.
        \label{eq:logit_sims}
    \end{equation}

The values of the coefficients were:
$b_{z} =0.87 $; $b_{0} = -0.96$; $b_{1} =2.9 $ and $b_{2} =-0.086 $.

The parameters of the models in this simulation study are inspired by the variables in the Viana do Castelo study. Specifically, $Y$ plays the role of the IOTF in 2006 ($Y=1$ overweight; $Y=0$ normal weight); $Z$ stands for the gender ($Z=1$ male; $Z=0$ female); $Z_1$ represents the IOTF in 2000; and $Z_2$ is representative for the physical performance in ABD test. The choice of the simulated distributions for these variables was driven by what was observed in our case study.

\vspace{0.5cm}
\textbf{Missing data}
\vspace{0.25cm}

Let be $R=1$ when $\{Y,Z,Z_{1},Z_{2}\}$ is observed. We simulated a MAR and MCAR scenarios, in which $R$ only depends on $Z$.
So, given $Z=z$, $1-R$ was drawn from $Ber(p_{4}(z))$, where $p_{4}(z)$ represented the proportion of missing data. We consider four different scenarios for $p_4(z)$.
The scenario S1, where $p_{4}(z) = 0.09z + 0.09(1 -z) $ that represents the $9\%$ of balanced missing values in the original data set. In scenario S2 there are $20\%$ of balanced  missing values: $p_{4}(z) = 0.20z + 0.20(1 -z) $. In the scenario S3, where $p_{4}(z) = 0.30z + 0.10(1 -z) $ there are $20\%$ unbalanced missing values, and the scenario S4 represents $35\%$ unbalanced missing values $p_{4}(z) = 0.65z + 0.05(1 -z) $. The balanced scenarios, S1 and S2 corresponded to MCAR context and unbalanced scenarios, S3 and S4 correspond to MAR context.

In our simulation design, when $R = 0$ then $\{Y,Z_{1},Z_{2}\}$ is missing  and only $Z$ is observed. On the other hand, $\{Y,Z,Z_{1},Z_{2}\}$ is observed when $R=1$. This type of blockwise missing data is the one occurring in the Viana do Castelo study.

\subsection{Estimates}

We used the following methods to estimate the regression coefficients $b_{0}$, $b_{z}$, $b_{1}$ and $b_{2}$:
\begin{itemize}
    \item (C) Model estimation with complete sample size (unrealistic; gold standard): Fit a logistic model to $Y$ depending on $Z$, $Z_{1}$ and $Z_{2}$.
    Obtain the estimates of the regression coefficients, $b_{0c}$, $b_{zc}$, $b_{1c}$ and $b_{2c}$. 
    \vspace{0.25cm}
   
    \item (CC) Complete case analysis: Fit a logistic model to $Y$ depending on $Z$, $Z_{1}$ and $Z_{2}$ considering the
    observations $\{Y_{i},Z_{i},Z_{1i},Z_{2i}\}_{i=1}^{n}$ with no missing values, that is, the observations with $R_{i} = 1$. Obtain the estimates of the regression coefficients, $b_{0cc}$, $b_{zcc}$, $b_{1cc}$ and $b_{2cc}$.
    \vspace{0.25cm}
    
    \item (IPW) Inverse probability weighting: Fit a logistic model to $Y$ depending on $Z$, $Z_{1}$ and $Z_{2}$ considering the
    observations $\{Y_{i},Z_{i},Z_{1i},Z_{2i}\}_{i=1}^{n}$ with no missing values and weighting
    each no missing observation by $1/\pi_{i}$, where $\pi_{i} = E(R_{i}|Y_{i},Z_{i},Z_{1i},Z_{2i})$.
    \vspace{0.2cm}
    \begin{itemize}
    \item IPW1: Use the theoretical value of $\pi_{i}$ calculated from $p_{4}(z)$.
    \vspace{0.15cm}
    \item IPW2: Estimate $\pi_{i}$ by applying a logistic model to $R$ depending on $Z$, under MAR assumption.

    \end{itemize}
    
    Obtain the estimates of the regression coefficients $b_{0ipw}$, $b_{zipw}$; $b_{1ipw}$ and $b_{2ipw}$.
    \vspace{0.25cm}

    \item (MI) Multiple imputation: For each imputed copy of the data, a logistic model to $Y$ depending on
    $Z$, $Z_{1}$ and $Z_{2}$ was considered and the estimates were combined into an overall estimate by using Rubin's rules (pool function).
    We obtained the estimates of the regression coefficients $b_{0mi}$, $b_{zmi}$, $b_{1mi}$ and $b_{2mi}$. The number of imputations was 5 (MI5) or 20 (MI20).
    \vspace{0.25cm}

    \item (ML) Maximum likelihood: We obtained the estimates of the regression coefficients $b_{0ml}$, $b_{zml}$, $b_{1ml}$ and $b_{2ml}$.
\end{itemize}

In order to evaluate CC and IPW estimates we used \texttt{glm} function of \texttt{R} with binomial family \cite{stats}. For MI we used  \texttt{mice} function of \texttt{mice} library in \texttt{R} with $m = 5$ and $m = 20$, where $m$ is the number of imputed data sets \cite{JSSv045i03}. In ML we used the function \texttt{frm\_em} of library \texttt{mdmb} in \texttt{R} that estimate the model with numerical integration \cite{mdmb}. In MI and ML we used the methods \texttt{"norm"} for $Z_{2}$ and \texttt{"logreg"} for $Z_{1}$ and $Y$.

Observe that in IPW we estimate $\pi_{i}$ with a logistic rule, which favours this approach because the missing model is logistic and we also use a logistic model for the weights. In MI and ML we also use the distributions with which we generate the data, so all methods are in favorable conditions.

Based on $M=1000$ trials we estimated the bias, the variance and the mean square error (MSE) of the estimators. Specifically, for each regression coefficient $b$ we calculated
\begin{equation*}
    Bias(\hat{b}_{method}) =\frac{1}{M} \sum_{k=1}^{M} \hat{b}_{k;method}-b= \hat E(\hat{b}_{k;method})-b
\end{equation*}
\begin{equation*}
    Var(\hat{b}_{method}) =\frac{1}{M} \sum_{k=1}^{M} [\hat{b}_{k;method}-\hat E(\hat{b}_{k;method})]^{2}
\end{equation*}
\begin{equation*}
   MSE(\hat{b}_{method}) =Bias(\hat{b}_{method})^{2}+Var(\hat{b}_{method})
\end{equation*}
where $\hat{b}_{method}$ denotes the generic estimator $\hat{b}$ of $b_{0}$, $b_{z}$, $b_{1}$ or $b_{2}$ based on the method C, CC, IPW, MI, ML, respectively, and $\hat{b}_{k,method}$ denotes the same estimator $\hat{b}$ when computed from the $k$-th Monte Carlo trial.

\subsection{Results}

The results for scenario S1 are reported in Table \ref{fig:S1}. From these Table it is seen that the bias, the variance and the MSE decrease as the sample size $n$ increases, indicating the consistency of the methods.
In this scenario S1, for any of the methods and sample sizes, bias has little influence on the MSE.
As expected, method C is the one reporting the smallest MSE, except when $n=230$. 
Among the realistic methods, MI and ML are the methods that gives the MSE closest to zero. For $n=230$, to $b_{0}$ and $b_{1}$, the ML present the MSE closest to zero and MI5 present MSE closest to zero to $b_{z}$. To $b_{2}$ all methods present the same MSE in any dimension.
For $n=400$ the MI20 present the MSE closest to zero for all estimators. For $n=1000$ ML and MI20 are similar to almost all estimators and present the MSE closest to zero. The results of CC method and IPW1 are similar because S1 is a MCAR context so is expected that the results of these methods are the same. In this scenario the methods that present higest MSE are CC and IPW.

In summary for $n=230$ the best model is ML to $b_{0}$ and $b_{1}$, to $b_{z}$ is MI. For $n=400$ the best method is MI20. For $n=1000$ is ML, with MI20 equaled in several estimates. Regardless of size to $b_{2}$ the MSE is equal to all methods. IPW is very similar to CC and often is the worst method.

In most cases, the MSE of MI20 was lower than the respective MSE of MI5. This agrees with the known result that more imputations tend to give better results. However, to $b_{z}$ for $n=230$ MI5 has lower MSE than MI20. Some unusual cases like this are cited in the literature, for instance in \cite{Graham} happened that MI with 5 imputations presented lower MSE than MI with 10 imputations.

\begin{table}[!ht] \tiny
\caption {Bias, Variance and Mean Square Error (MSQ) to estimated coefficients for the logistic regression model provided by five different methods: complete case analysis (CC), inverse probability weighting (IPW), multiple imputation with 5 and 20 imputations (MI5, MI20), and maximum likelihood (ML) of scenario S1 (9\% balanced missing data) from different n}
\centering
\begin{tabular}{|c||c|c|c||c|c|c||c|c|c||c|c|c|}
\hline
S1 & & $b_{0}$ & & & $b_{z}$ & & & $b_{1}$ & & & $b_{2}$ & \cr 
$n=230$ & Bias & Var & MSE & Bias & Var & MSE & Bias & Var & MSE & Bias & Var & MSE\\			
\hline
C&	-0.0179& 1.2213& 1.2216	&	0.0552& 0.2263& 0.2293&		0.0989& 0.5730 &0.5828	& -0.0041 &0.0009& \textbf{0.0010}\\
CC	&-0.0289& 1.3402& 1.3410&		0.0581 &0.2560 &0.2594&		0.1129 &0.6217& 0.6344&	 -0.0043& 0.0010& \textbf{0.0010}\\
IPW1&	-0.0289& 1.3417& 1.3425&		0.0581 &0.2560 &0.2594&		0.1130& 0.6230& 0.6358&	 -0.0043& 0.0010& \textbf{0.0010}\\
IPW2&	-0.0293& 1.3415& 1.3424&		0.0580 &0.2560& 0.2594&		0.1129& 0.6232& 0.6359&	 -0.0043& 0.0010 &\textbf{0.0010}\\
MI5&	-0.0081& 1.1487& 1.1488&		0.0536& 0.2527& \textbf{0.2556}&		0.0896& 0.4393& 0.4473&	 -0.0040& 0.0010& \textbf{0.0010}\\
MI20&	-0.0134& 1.3086& 1.3088&		0.0547& 0.2537 &0.2567&		0.1000& 0.5889 &0.5989&	 -0.0042& 0.0010& \textbf{0.0010}\\
ML&	-0.0054& 1.0294& \textbf{1.0294}	&	0.0571& 0.2557& 0.2590&		0.0973 &0.3428 &\textbf{0.3523}	& -0.0046 &0.0010 &\textbf{0.0010}\\
\hline

\hline
S1 & & $b_{0}$ & & & $b_{z}$ & & & $b_{1}$ & & & $b_{2}$ & \cr 
$n=400$ & Bias & Var & MSE & Bias & Var & MSE & Bias & Var & MSE & Bias & Var & MSE\\
\hline
C	&-0.0471 &0.4735& 0.4757&	 	0.0261& 0.1370& 0.1377&	 	0.0674& 0.1314& 0.1359&	 	-0.0010&	0.0005 &0.0005\\
CC	&-0.0414& 0.5360& 0.5377	&	0.0281& 0.1557& 0.1565&	 	0.0760& 0.1513& 0.1571&	 	-0.0015& 0.0006& \textbf{0.0006}\\
IPW1	&-0.0414& 0.5360 &0.5377	&	0.0281 &0.1557& 0.1565&		0.0760& 0.1513& 0.1571	&	-0.0015& 0.0006 &\textbf{0.0006}\\
IPW2&	-0.0418 &0.5349 &0.5366	& 	0.0280& 0.1557 &0.1565&		0.0761& 0.1514& 0.1572&		-0.0015 &0.0006 &\textbf{0.0006}\\
MI5	&-0.0352& 0.5372& 0.5384&		0.0239& 0.1563& 0.1569&	 	0.0688& 0.1527& 0.1574	& 	-0.0013& 0.0006 &\textbf{0.0006}\\
MI20&	-0.0311 &0.5345& \textbf{0.5355}	&	0.0254& 0.1553& \textbf{0.1559}	&	0.0683& 0.1496 &\textbf{0.1543}	&	-0.0015& 0.0006& \textbf{0.0006}\\
ML	&-0.0303& 0.5351 &0.5360	& 	0.0272& 0.1556& 0.1563&		0.0727 &0.1508& 0.1561	&	-0.0018& 0.0006 &\textbf{0.0006}\\
\hline

\hline
S1 & & $b_{0}$ & & & $b_{z}$ & & & $b_{1}$ & & & $b_{2}$ & \cr 
$n=1000$ & Bias & Var & MSE & Bias & Var & MSE & Bias & Var & MSE & Bias & Var & MSE\\
\hline
C	&-0.0147& 0.1901& 0.1903&		0.0074& 0.0514 &0.0515 	&	0.0285& 0.0571& 0.0579 	&	-0.0006& 2e-04& 2e-04\\
CC	&-0.0160& 0.2137& 0.2140&		0.0107& 0.0568 &\textbf{0.0569} &		0.0323& 0.0625& 0.0635 &		-0.0007& 2e-04& \textbf{2e-04}\\
IPW1&	-0.0160& 0.2137& 0.2140&		0.0107& 0.0568& \textbf{0.0569}&		0.0323& 0.0625& 0.0635 	&	-0.0007 &2e-04 &\textbf{2e-04}\\
IPW2&	-0.0158& 0.2136& 0.2138&		0.0107& 0.0568& \textbf{0.0569}&		0.0322& 0.0625& 0.0635	&	-0.0007 &2e-04 &\textbf{2e-04}\\
MI5&	-0.0105& 0.2185& 0.2186&		0.0102& 0.0584 &0.0585	&	0.0288 &0.0632& 0.0640	&	-0.0008 &2e-04& \textbf{2e-04}\\
MI20&	-0.0122& 0.2140& 0.2141&		0.0101 &0.0568 &\textbf{0.0569}&		0.0286 &0.0624 &\textbf{0.0632}&		-0.0007& 2e-04 &\textbf{2e-04}\\
ML&	-0.0051& 0.2135& \textbf{0.2135}	&	0.0098& 0.0568& \textbf{0.0569}	&	0.0291 &0.0624& \textbf{0.0632} 	&	-0.0010& 2e-04& \textbf{2e-04}\\
\hline
\end{tabular}
 \label{fig:S1}
\end{table}

Similar results to those of S1 were obtained for scenarios S2 (Table \ref{fig:S2}), S3 (Table \ref{fig:S3}) and S4 (Table \ref{fig:S4}) in terms of decrease of bias, variance and MSE with $n$ increase. Also the contribution of the bias to the MSE was negligible and the gold standard C provided the best results in all the cases.

In the second scenario S2 (20\% balanced missing data) CC and IPW reported similar results, like S1. Table \ref{fig:S2} 
shows that, for $n=230$, the second method with MSE closest to zero (the first method is C) to estimate $b_{0}$ and $b_{1}$ is ML, while for $b_{z}$ the results closest to zero are provided by MI20. For $n=400$ MI20 is a methodology that present MSE closest to zero in all parameters.  
For $n=1000$ MI20 present the MSE closest to zero in $b_{z}$ and $b_{1}$, while in $b_{0}$ is the ML method. 
To estimate $b_{2}$, in all sizes, all methods provide the same MSE.
Note that for $n=230$ the IPW and CC are the worst models, with big differences in MSE, especially in $b_{0}$ and $b_{1}$, for $n=400$ and $n=1000$ the differences between the methods are much smaller.
However IPW is the method that present highest MSE in $n=230$ and $n=400$. For $n=1000$ is MI5 that present highest MSE in $b_{0}$ and $b_{z}$ and CC and IPW present highest MSE to $b_{1}$.

In summary, for $n=230$ the best method is ML to $b_{0}$ and $b_{1}$ and MI20 to $b_{z}$. For the other dimensions MI20 is the best method, except to $b_{0}$ for $n=1000$. The method with worst results is IPW for $n=230$ and $n=400$.
\medskip

\begin{table}[!ht] \tiny
\caption {Bias, Variance and Mean Square Error (MSQ) to estimated coefficients for the logistic regression model provided by five different methods: complete case analysis (CC), inverse probability weighting (IPW), multiple imputation with 5 and 20 imputations (MI5, MI20), and maximum likelihood (ML) of scenario S2 (20\% balanced missing data) from different n}
\centering
\begin{tabular}{|c||c|c|c||c|c|c||c|c|c||c|c|c|}
\hline
S2 & & $b_{0}$ & & & $b_{z}$ & & & $b_{1}$ & & & $b_{2}$ & \cr 
$n=230$ & Bias & Var & MSE & Bias & Var & MSE & Bias & Var & MSE & Bias & Var & MSE\\
\hline
C	&-0.0179& 1.2213 &1.2216	&	0.0552& 0.2263 &0.2293	&	0.0989& 0.5730& 0.5828&	 -0.0041& 0.0009& 0.0009\\
CC	&-0.0643& 2.3852& 2.3893	&	0.0605& 0.3094& 0.3131&		0.1636& 1.5632& 1.5900&	 -0.0051& 0.0012& \textbf{0.0012}\\
IPW1&	-0.0647& 2.3980& 2.4022&		0.0605& 0.3094 &0.3131&		0.1640& 1.5759 &1.6028&	 -0.0051& 0.0012& \textbf{0.0012}\\
IPW2&	-0.0647& 2.3712 &2.3754&		0.0603 &0.3095& 0.3131&		0.1632& 1.5410 &1.5676&	 -0.0051& 0.0012& \textbf{0.0012}\\
MI5	& 0.0075& 1.5473& 1.5474	&	0.0491& 0.3065 &0.3089&		0.0937 &0.7297 &0.7385&	 -0.0047 &0.0012& \textbf{0.0012}\\
MI20&	 0.0022& 1.6203& 1.6203&		0.0443& 0.2978 &\textbf{0.2998}&		0.1016& 0.7897& 0.8000&	 -0.0045& 0.0012& \textbf{0.0012}\\
ML	&-0.0041 &1.2845& \textbf{1.2845}&		0.0593& 0.3089 &0.3124&		0.1120 &0.4630 &\textbf{0.4755}&	 -0.0054 &0.0012 &\textbf{0.0012}\\
\hline

\hline
S2 & & $b_{0}$ & & & $b_{z}$ & & & $b_{1}$ & & & $b_{2}$ & \cr 
$n=400$ & Bias & Var & MSE & Bias & Var & MSE & Bias & Var & MSE & Bias & Var & MSE\\
\hline
C	&-0.0471 &0.4735& 0.4757	& 	0.0261& 0.1370 &0.1377&	 	0.0674 &0.1314& 0.1359&	 	-0.0010&	0.0005 &0.0005\\
CC	&-0.0532 &0.6164 &0.6192	&	0.0297 &0.1802& 0.1811&		0.0728& 0.1639& 0.1692&		-0.0014 &0.0006& \textbf{0.0006}\\
IPW1&	-0.0532 &0.6164& 0.6192	&	0.0297 &0.1802& 0.1811&		0.0728 &0.1639& 0.1692&		-0.0014& 0.0006 &\textbf{0.0006}\\
IPW2&	-0.0533& 0.6184& 0.6212&		0.0297& 0.1804& 0.1813&		0.0728& 0.1643& 0.1696&		-0.0013& 0.0006& \textbf{0.0006}\\
MI5	&-0.0227 &0.6220& 0.6225	&	0.0174 &0.1806& 0.1809	&	0.0520&0.1629 &0.1656&		-0.0012& 0.0006& \textbf{0.0006}\\
MI20&	-0.0270& 0.6128 &\textbf{0.6135}&		0.0234 &0.1792 &\textbf{0.1797}		&0.0531 &0.1595&\textbf{0.1623}		&-0.0012 &0.0006 & \textbf{0.0006}\\
ML&	-0.0404 &0.6163& 0.6179&		0.0287 &0.1801 &0.1809&		0.0688 &0.1633& 0.1680&		-0.0017& 0.0006 &\textbf{0.0006}\\
\hline

\hline
S2 & & $b_{0}$ & & & $b_{z}$ & & & $b_{1}$ & & & $b_{2}$ & \cr 
$n=1000$ & Bias & Var & MSE & Bias & Var & MSE & Bias & Var & MSE & Bias & Var & MSE\\
\hline
C&	-0.0147& 0.1901& 0.1903&		0.0074 &0.0514& 0.0515 &		0.0285 &0.0571 &0.0579 	&	-0.0006 &2e-04 &\textbf{2e-04}\\
CC&	-0.0195& 0.2496& 0.2500&		0.0048& 0.0615& 0.0615&		0.0378& 0.0752& 0.0766&		-0.0007 &2e-04 &\textbf{2e-04}\\
IPW1&	-0.0195& 0.2496 & 0.2500&		0.0048& 0.0615& 0.0615&		0.0378 &0.0752& 0.0766&		-0.0007 &2e-04 &\textbf{2e-04}\\
IPW2&	-0.0194& 0.2494& 0.2498&		0.0048 &0.0615& 0.0615&		0.0378 &0.0752 &0.0766&		-0.0007 &2e-04 &\textbf{2e-04}\\
MI5&	-0.0115& 0.2543& 0.2544&		0.0026& 0.0640& 0.0640&		0.0312 &0.0755 &0.0765& 		-0.0006 &2e-04 &\textbf{2e-04}\\
MI20&	-0.0113& 0.2510& 0.2511&		0.0010 &0.0613& \textbf{0.0613}&		0.0294 &0.0750 &\textbf{0.0759}&		-0.0005& 2e-04 &\textbf{2e-04}\\
ML&	-0.0073 &0.2493& \textbf{0.2494}&		0.0038 &0.0615 &0.0615	&	0.0340 &0.0750& 0.0762	&	-0.0010& 2e-04 &\textbf{2e-04}\\
\hline

\end{tabular}
 \label{fig:S2}
\end{table}

In scenario S3 (20\% of unbalanced missing data), see Table \ref{fig:S3}, for $n=230$
the values of MSE closest to zero correspond to ML for the coefficients $b_{0}$ and $b_{1}$. For the $b_{z}$ coefficient, the IPW method presents results closest to zero.
However, to estimate $b_{2}$ CC, MI20 and ML have the same MSE that is the lowest. In this dimension MI20 present MSE higher that MI5 to $b_{0}$ and $b_{1}$. In this case, IPW2 present higher MSE to $b_{0}$ and $b_{1}$. For $b_{z}$ is MI5 that present the highest MSE.
For $n=400$ MSE is closest to zero in ML for all coefficients. Although, for $b_{2}$ the MSE is equal between all methods. In this dimension MI5 is the method that present higher MSE to $b_{0}$ and $b_{z}$, while to $b_{1}$ the method is IPW. 
For $n=1000$ MSE is closest to zero in ML to all coefficients, while MI5  is the method with MSE further away from zero. For $n=400$ and $n=1000$ the MSE to $b_{2}$ is equal between all methods, except MI5 for $n=1000$.

In summary ML often has the lowest MSE for all $n$ and all betas, overcoming MI.
IPW behaves very similar to ML for $b_{z}$, but it has very high MSE for $b_{0}$ and $b_{1}$ with $n=230$.
In some situations, MI20's MSE are worse than MI5's.

In the last scenario S4 (35\% imbalanced missing data), Table \ref{fig:S4}, MSE results are closest to zero in MI20 method to estimate all coefficients in all dimensions, except to $b_{1}$ in $n=230$, where the MSE closest to zero is in ML. In this scenario IPW2 is the method that present the highest MSE. 

In summary, MI20 is the best method, except for $n=230$ to $b_{1}$ and the worst method is IPW.

\begin{table}[!ht] \tiny
\caption {Bias, Variance and Mean Square Error (MSQ) to estimated coefficients for the logistic regression model provided by five different methods: complete case analysis (CC), inverse probability weighting (IPW), multiple imputation with 5 and 20 imputations (MI5, MI20), and maximum likelihood (ML) of scenario S3 (20\% unbalanced missing data) from different n}
\centering
\begin{tabular}{|c||c|c|c||c|c|c||c|c|c||c|c|c|}
\hline
S3 & & $b_{0}$ & & & $b_{z}$ & & & $b_{1}$ & & & $b_{2}$ & \cr 
$n=230$ & Bias & Var & MSE & Bias & Var & MSE & Bias & Var & MSE & Bias & Var & MSE\\
\hline
C	&-0.0179 &1.2213& 1.2216&		0.0552& 0.2263& 0.2293&		0.0989& 0.5730 &0.5828	& -0.0041& 0.0009& 0.0009\\
CC&	-0.1090 &2.9393& 2.9512&		0.0684 &0.2934 &0.2981	&	0.2154& 2.1945& 2.2409	& -0.0051 &0.0012& \textbf{0.0012}\\
IPW1&	-0.1170& 3.0003& 3.0140	&	0.0680& 0.2930& \textbf{0.2976}&		0.2160& 2.2342& 2.2809	& -0.0049 &0.0013& 0.0013\\
IPW2&	-0.1168& 3.0050& 3.0186&		0.0681 &0.2932& 0.2978&		0.2165 &2.2345& 2.2814&	 -0.0049& 0.0013& 0.0013\\
MI5	&-0.0688& 1.9804& 1.9851&		0.0864& 0.2915& 0.2990	&	0.1433 &1.2009& 1.2214 &	 -0.0041 &0.0013& 0.0013\\
MI20&	-0.1090& 2.9393& 2.9512&		0.0684& 0.2934 &0.2981&		0.2154 &2.1945 &2.2409&	 -0.0051 &0.0012& \textbf{0.0012}\\
ML	&-0.0250 &1.3554 &\textbf{1.3560}	&	0.0680 &0.2931 &0.2977&		0.1398& 0.5691& \textbf{0.5886}	& -0.0054 &0.0012 &\textbf{0.0012}\\
\hline

\hline
S3 & & $b_{0}$ & & & $b_{z}$ & & & $b_{1}$ & & & $b_{2}$ & \cr 
$n=400$ & Bias & Var & MSE & Bias & Var & MSE & Bias & Var & MSE & Bias & Var & MSE\\
\hline
C	&-0.0471 &0.4735& 0.4757&	 	0.0261& 0.1370 &0.1377&	 	0.0674& 0.1314& 0.1359&	 	-0.0010&	0.0005& 0.0005\\
CC&	-0.0453 &0.6351& 0.6372&		0.0321& 0.1757 &0.1767&		0.0785 &0.1858& 0.1920&		-0.0017 &0.0007 &\textbf{0.0007}\\
IPW1&	-0.0490& 0.6424& 0.6448&		0.0313& 0.1747& \textbf{0.1757}&		0.0781& 0.1903& 0.1964&		-0.0016& 0.0007 &\textbf{0.0007}\\
IPW2&	-0.0493& 0.6431& 0.6455&		0.0314& 0.1747& \textbf{0.1757}&		0.0785& 0.1902& 0.1964&		-0.0016 &0.0007& \textbf{0.0007}\\
MI5 &	-0.0252& 0.6552& 0.6558&		0.0422& 0.1795& 0.1813&		0.0575& 0.1880& 0.1913& 		-0.0016& 0.0007& \textbf{0.0007}\\
MI20&	-0.0453& 0.6351& 0.6372&		0.0321& 0.1757& 0.1767&		0.0785& 0.1858& 0.1920&		-0.0017& 0.0007& \textbf{0.0007}\\
ML&	-0.0322& 0.6336& \textbf{0.6346}	&	0.0317& 0.1756 & 0.1766	&	0.0742 &0.1851 &\textbf{0.1906}		&-0.0020& 0.0007 &\textbf{0.0007}\\
\hline

\hline
S3 & & $b_{0}$ & & & $b_{z}$ & & & $b_{1}$ & & & $b_{2}$ & \cr 
$n=1000$ & Bias & Var & MSE & Bias & Var & MSE & Bias & Var & MSE & Bias & Var & MSE\\
\hline
C	&-0.0147 &0.1901& 0.1903&		0.0074& 0.0514& 0.0515& 		0.0285& 0.0571& 0.0579 	&	-0.0006& 2e-04 &2e-04\\
CC&	-0.0186& 0.2425& 0.2428	& 	0.0092 &0.0628& 0.0629&		0.0390& 0.0761& 0.0776	&	-0.0009 &2e-04 &\textbf{2e-04}\\
IPW1&	-0.0176& 0.2469& 0.2472&	 	0.0090& 0.0627& \textbf{0.0628}	&	0.0378 &0.0775& 0.0789 &		-0.0009 &2e-04 &\textbf{2e-04}\\
IPW2&	-0.0175& 0.2466 &0.2469&		0.0091& 0.0627 & \textbf{0.0628} &		0.0378& 0.0776& 0.0790 &		-0.0009 &2e-04 &\textbf{2e-04}\\
MI5&	-0.0092& 0.2493& 0.2494&	 	0.0143 &0.0650& 0.0652&		0.0293 &0.0790& 0.0799	&	-0.0009 &3e-04 &3e-04\\
MI20&	-0.0186& 0.2425& 0.2428&		0.0092 &0.0628& 0.0629&		0.0390& 0.0761 &0.0776&		-0.0009 &2e-04 &\textbf{2e-04}\\
ML	&-0.0060 &0.2419 &\textbf{0.2419}&		0.0086& 0.0627 &\textbf{0.0628}	&	0.0348& 0.0758 &\textbf{0.0770}		&-0.0012& 2e-04& \textbf{2e-04}\\
\hline

\end{tabular}
\label{fig:S3}
\end{table}

\begin{table}[!ht] \tiny
\caption {Bias, Variance and Mean Square Error (MSQ) to estimated coefficients for the logistic regression model provided by five different methods: complete case analysis (CC), inverse probability weighting (IPW), multiple imputation with 5 and 20 imputations (MI5, MI20), and maximum likelihood (ML) of scenario S4 (35\% unbalanced missing data) from different n}
\centering
\begin{tabular}{|c||c|c|c||c|c|c||c|c|c||c|c|c|}
\hline
S4 & & $b_{0}$ & & & $b_{z}$ & & & $b_{1}$ & & & $b_{2}$ & \cr 
$n=230$ & Bias & Var & MSE & Bias & Var & MSE & Bias & Var & MSE & Bias & Var & MSE\\
\hline
C&	-0.0179 &1.2213 &1.2216&		0.0552& 0.2263& 0.2293	&	0.0989 &0.5730 & 0.5828&	-0.0041& 0.0009& 0.0009\\
CC&	-0.2499 &6.6674& 6.7299&		0.0583& 0.4782& 0.4816	&	0.4437 &5.4602&  5.6571	&-0.0082 &0.0017 & 0.0018\\
IPW1&	-0.2974& 7.2612& 7.3496&		0.0661 &0.5002& 0.5046&		0.4892& 5.7872 & 6.0265	&-0.0082 &0.0022& 0.0023\\
IPW2&	-0.2997& 7.2891 &7.3789&		0.0660& 0.5033 &0.5077&		0.4915& 5.7931 & 6.0347	&-0.0082& 0.0022 &0.0023\\
MI5	&-0.0083& 1.8805 &1.8806	&	0.0414& 0.3105 &0.3122	&	0.1111 &0.9962& 1.0085	&-0.0044 &0.0012& \textbf{0.0012}\\
MI20&	 0.0004 &1.7769& \textbf{1.7769}	&	0.0463 &0.3008& \textbf{0.3029}&		0.1058& 0.9546& 0.9658&	-0.0047& 0.0012 &\textbf{0.0012}\\
ML&	-0.0266& 1.9582& 1.9589		&0.0583 &0.4758& 0.4792&		0.2278& 0.8590& \textbf{0.9109}&	-0.0085& 0.0017& 0.0018\\
\hline

\hline
S4 & & $b_{0}$ & & & $b_{z}$ & & & $b_{1}$ & & & $b_{2}$ & \cr 
$n=400$ & Bias & Var & MSE & Bias & Var & MSE & Bias & Var & MSE & Bias & Var & MSE\\
\hline
C &	 	-0.0471 &0.4735 &0.4757&0.0261& 0.1370 &0.1377&	 	0.0674 &0.1314& 0.1359&	 	-0.0010&	0.0005& 0.0005\\
CC&	-0.0574& 0.8687& 0.8720&		0.0288& 0.2580& 0.2588&		0.1077& 0.2634& 0.2750&		-0.0024 &0.0010& 0.0010\\
IPW1& 	-0.0855& 1.1125& 1.1198 &		0.0295& 0.2623 &0.2632&		0.1270& 0.3470& 0.3631&		-0.0021& 0.0012 &0.0012\\
IPW2&	-0.0864 &1.1191 &1.1266& 		0.0296& 0.2630& 0.2639&		0.1281 &0.3489 &0.3653&		-0.0021 &0.0012& 0.0012\\
MI5&	-0.0274 &0.6425 &0.6433	&	0.0199& 0.1819& 0.1823&		0.0497& 0.1682 &0.1707	&	-0.0010& 0.0006& \textbf{0.0006}\\
MI20&	-0.0284& 0.6108& \textbf{0.6116}&		0.0256 &0.1781& \textbf{0.1788}	&	0.0507& 0.1593 &\textbf{0.1619}		&-0.0012& 0.0006 &\textbf{0.0006}\\
ML&	-0.0437& 0.8618& 0.8637	&	0.0290& 0.2570 &0.2578	&	0.1021& 0.2611& 0.2715		&-0.0027& 0.0010 &0.0010\\
\hline

\hline
S4 & & $b_{0}$ & & & $b_{z}$ & & & $b_{1}$ & & & $b_{2}$ & \cr 
$n=1000$ & Bias & Var & MSE & Bias & Var & MSE & Bias & Var & MSE & Bias & Var & MSE\\
\hline
C&	-0.0147& 0.1901 &0.1903	&	0.0074& 0.0514& 0.0515& 		0.0285&0.0571 &0.0579& 		-0.0006& 2e-04 &2e-04\\
CC	&-0.0176 &0.3290 &0.3293&		0.0157& 0.0957& 0.0959&		0.0565 &0.1025& 0.1057&		-0.0015 &3e-04 &3e-04\\
IPW1&	-0.0133 &0.4072 &0.4074&		0.0192& 0.0986& 0.0990 &		0.0630 &0.1283& 0.1323 	&	-0.0020 &4e-04& 4e-04\\
IPW2&	-0.0140& 0.4083& 0.4085&		0.0193& 0.0987& 0.0991&		0.0635& 0.1291 &0.1331	&	-0.0020 &4e-04 &4e-04\\
MI5&	-0.0118 &0.2607& 0.2608	&	0.0020 &0.0627 &0.0627	&	0.0285 &0.0767& 0.0775 &	-0.0004& 2e-04& \textbf{2e-04}\\
MI20&	-0.0096 &0.2486 &\textbf{0.2487}	&	0.0014& 0.0620 &\textbf{0.0620}	&	0.0288& 0.0756& \textbf{0.0764}	&	-0.0006 &2e-04 &\textbf{2e-04}\\
ML	&-0.0043& 0.3271& 0.3271&		0.0160 &0.0953 &0.0956&		0.0515& 0.1019 &0.1046&		-0.0019 &3e-04 &3e-04\\
\hline

\end{tabular}
\label{fig:S4}
\end{table}

In general, it can be said that all the methods work consistently. Furthermore, the conclusions about the best method are very similar for the simulated scenarios. In general, the MI and ML methodology are the results with MSE closest to zero. 
The scenario S1, that is in a MCAR context with small proportion of missing, the MI or ML present the MSE results closest to zero according de dimension of sample.  MI for $n=400$ and ML for $n=1000$. 
In the other scenario MCAR, S2, with 20\% of proportion of missing, the method with smaller MSE for $n=400$ is MI. In the others dimensions MI and ML are the methods that present smaller MSE depending on the dimension and on the parameter.  The balanced scenarios present smaller MSE with MI or ML.
IPW and CC are the methods that present biggest MSE in MCAR context, except in S2 with big sample.

In case of scenario S3, MAR context with unbalanced proportion of missing, the parameter $b_{z}$ present MSE closest to zero with IPW method, while the others parameters present smaller MSE with ML. In other hand, the method that present biggest MSE is MI to big sample. To the others dimensions the worst methods are IPW or MI depending on the parameter and sample size.

In case of MAR context with unbalanced and big proportion of missing, scenario S4, MSE is closest to zero with the MI methodology. In other hand MSE is biggest with IPW.

In general, S1 and S2 present lower MSE with ML and MI. ML is the method with lowest MSE in S3 and in S4 is the MI method that present lowest MSE.

From these results it can be concluded that, when the missing proportion is balanced, the best method is MI or ML. When the missing proportion is unbalanced, the best method is MI if the proportion of missing is big or ML if the proportion of missing is not so big.

The definitive conclusions on the relative performance of the methods to handle missing data is aligned with the literature. Carpenter \cite{carpenter2006comparison} 
stated the conditions under which CC is consistent and when it is not. In our study, the CC methodology was never presented with the best results and not always as the one with the worst results in terms of MSE. 
However complete case analysis is valid, that is, the method provides consistent estimators just in MCAR context or in MAR context if the probability of missing is independent of response variable. In our simulation study the distribution of $R$ does not depend on the outcome, and \cite{carpenter2021missing} highlights that this fact produces consistency of the CC estimators. 
On the other hand, our simulation results for IPW and CC methods are similar. This behavior is natural since, according to \cite{carpenter2021missing},  \textit{"IPW results will not differ markedly from the complete records unless the covariates are MAR given the dependent variable".} 
Although CC and IPW methods provide consistent estimators, if the sample size is not large enough (for example, $n=230$) in general these estimators are clearly less efficient than the ML estimators and those obtained by the MI method. In our simulation MI and ML provide the best results (smaller MSE) in balanced and unbalanced scenarios.

In fact the way to deal with missing data is not consensual. There are studies, like Mhike and Liu \cite{mhike2021, liu2019missing} that present MI as the best method; while others, like \cite{allison2012handling}, conclude that ML is the best. Peng \cite{peng2008} compared ML and MI methods for handling missing data in categorical covariates in logistic regression. In general, our results are not similar to those in that paper because the results of their simulation study favored MI over ML. On the other hand, Raghunathan \cite{Raghunathan} revised three approaches for analyzing incomplete data: IPW, MI and ML and used the same logistic regression example to compare them. His model considered a binary outcome, a binary covariate and a continuous covariate, with missing data only in the continuous variable. He concluded that the estimates based on MI are in general more efficient than the IPW estimates. In this study we concluded the same. Due to the lack of conclusive research on the relative performance of the methods, Jelivcic  \cite{jelivcic2010missing} suggests the simultaneous use of several of them and the evaluation of the concordance of the results. If the results are similar, the conclusions are assured.

To expand our study, the accuracy of the estimate for the standard error of the regression coefficients provided by each of the methods was evaluated. For C, CC and IPW, standard error corresponds to the standard error provided by \texttt{glm} function. On the other hand, for MI we consider the standard error of pool of fit model provided, according Rubin's rules, by \texttt{mice} function. Finally, in ML we consider the standard errors provide by \texttt{frm\_em}. For each estimator, we computed the mean of the ratio between the standard error of each method and the Monte Carlo approximation of the true standard error. Values close to 1 indicate a good estimate, while values greater than 1 or less than 1 indicate an excessive or underestimated estimate, respectively. The average ratios along $1,000$ Monte Carlo trials for the simulation scenario S2 are reported in Table \ref{fig:Erro_S2}; the other scenarios provided similar results (not shown).

From Table \ref{fig:Erro_S2} it is seen that the standard error is accurately estimated except for IPW method, for which some underestimation occurs. This could be occur because the standard errors implemented in \texttt{glm} are appropriate only for non-random weights.

\begin{table}[!ht] \tiny
\caption {Accuracy of standard error of scenario S2}
\centering
\begin{tabular}{|c||c|c|c||c|c|c||c|c|c||c|c|c|}
\hline
S2 & & $b_{0}$ & & & $b_{z}$ & & & $b_{1}$ & & & $b_{2}$ & \cr
n  & 230 & 400 & 1000 & 230 & 400 & 1000 & 230 & 400& 1000 & 230 & 400& 1000\\
\hline
C     &  2.1249 &1.0086& 0.9939    &     1.0191 &0.9746& 0.9938      &   2.5248 &1.0144& 0.9552&       0.9816& 0.9844& 1.0003\\
CC    &  5.3265& 0.9955& 0.9719    &     0.9864 &0.9559& 1.0172      &   6.1844 &1.0238& 0.9339&       0.9687& 0.9867& 0.9939\\
IPW1  &  4.9428 &0.8904& 0.8692    &     0.8823 &0.8549& 0.9098      &   5.7452 &0.9157& 0.8353&       0.8665& 0.8826& 0.8890\\
IPW2  &  4.3581& 0.8889& 0.8696    &     0.8808 &0.8542& 0.9095      &   5.0504 &0.9149& 0.8352&       0.8646& 0.8811& 0.8889\\
MI5  &   4.2383 &1.0081& 0.9833    &     1.0044 &0.9705& 1.0224      &   5.5778 &1.0457& 0.9501&       0.9769& 0.9951& 0.9911\\
MI20 &   4.4202 &1.0026& 0.9747    &     1.0101 &0.9625& 1.0229      &   5.7704 &1.0424& 0.9409&       0.9797& 0.9892& 0.9980\\
ML   &   0.9565& 0.9952& 0.9721    &     0.8666 &1.0251& 0.9345      &   0.9682 &0.9844& 0.9920&       0.9867& 0.9558& 1.0171\\
\hline
\end{tabular}
 \label{fig:Erro_S2}
\end{table}

As mentioned, the results on the standard error for scenarios S1, S3 and S4 were similar to those in Table \ref{fig:Erro_S2}. However, for S2 we found that the mean ratios of IPW method were much lower than in the other three scenarios. Indeed, for scenario S2 it is seen that the standard error attached to IPW method is negatively biased; the only exception occurred for $n=230$ and the parameters $b_{0}$ and $b_{1}$. This is somehow in contrast to \cite{carpenter2021missing}, who play down the importance of ignoring the randomness of the $p_i$'s. It would be interesting to investigate the impact of the sample size on the accuracy of the standard error further; this is relevant, for instance, whenever an asymptotic formula is used to approximate the standard deviation of the estimate.

\section{Viana do Castelo study}

In this section several methods for the analysis of missing data are applied to the Viana do Castelo case study. Specifically, we consider CC, IPW, MI and ML methods as described in Section 2. The response variable $Y$ is the IOTF in 2006. As covariates we consider the gender, and the IOTF and ABD (physical performance) in 2000. These three covariates are denoted by $Z$, $Z_{1}$ and $Z_{2}$, respectively.

The table \ref{tab:viana} presents the coefficient estimates of the logistic regression model, together with the corresponding corresponding standard errors.

\begin{table}[!ht] \small  %\tiny
\centering
\caption{Estimated coefficients (standard errors in brackets) for the logistic regression model provided by five different methods: complete case analysis (CC), inverse probability weighting (IPW), multiple imputation with 5 and 20 imputations (MI5, MI20), and maximum likelihood (ML). Viana do Castelo study}
\begin{tabular}{|c|c|c|c|c|}
\hline
$n=230$ &  $b_0$  & $b_z$  & $b_1$  & $b_2$ \\
Method & Intercept & Gender & IOTF & ABD\\
\hline
CC& -0.0825  & -0.8759. & 2.9606*** &-0.0861** \\
  & (1.0087) & (0.4765) & (0.4781) & (0.0304) \\
IPW & -0.0819  & -0.8759.  &2.9605*** &-0.0861** \\
 & (0.9774) & (0.4619) & (0.4631) & (0.0295) \\
 MI5 & -0.2656  & -0.9109.  & 2.9834***  & -0.0804** \\
 & (0.9840) & (0.4756)  & (0.4790) & (0.0303) \\

MI20&  -0.1090  &  -0.8947. & 2.9771*** &  -0.0852** \\
& (1.0116) & (0.4744) & (0.4807) & (0.0304) \\

ML & -0.0599&    -0.8830.  &   2.9613***    &  -0.0866**  \\
 & (1.0086)& (0.4767)& (0.4775)&(0.0304)\\
 \hline
\end{tabular}
\label{tab:viana}
\end{table}

From Table \ref{tab:viana} it is seen that $b_{z}$ is not significant at level $\alpha=0.05$, which means that the gender variable is not relevant for the model. On the contrary, the parameters $b_{1}$ and $b_{2}$, that correspond to the variables IOTF and ABD in year 2000, are significant. The results obtained are intuitive. The estimate of the parameter $b_{1}$ is positive, which means that the variable IOTF in 2000 has a positive impact on the response variable. On the other hand, the estimate of the parameter $b_{2}$ is negative, which means that this variable will have a negative impact on the response variable. In other words, an individual with high IOTF (being overweight or obese) and low ABD in year 2000 will tend to have a high IOTF value in 2006, that is, the individual is still overweight or obese.
The null effect of gender is not entirely surprising, since the influence of this variable could be partly captured by the past value of IOTF or by the physical performance (ABD in 2000), which is different between genders.

As expected from the simulation results, the parameter estimates obtained with IPW and CC methods are similar to each other. The estimates provided by MI5, MI20 and ML are slightly different. All methods provide a correct classification of IOTF levels in more than 85\% of cases.
Checking the standard error associated with each of the estimators, it is seen that the estimators obtained by IPW are those with the lowest dispersion, although the differences among the several methods to this regard are small.

\section{Main conclusions}

In this paper we have investigated through simulations the relative performance of several methods for handling missing data in the scope of logistic regression. The simulation design was inspired by the Viana do Castelo study, in which obesity is predicted from its past value and the physical performance, controlling for gender too. Specifically, four different methods were compared: complete case analysis, inverse probability weighting, multiple imputation and maximum likelihood. All the methods exhibited a consistent behavior in the simulated scenarios; however, no method was uniformly the best. This is in well agreement with the literature on missing data.

Missing values in both the covariates and the response variable have been considered, which brings new insights to the literature on missing data.

In the simulations, MI and ML were the best methods because they presented a lower MSE. In both balanced and unbalanced scenarios with a small proportion of missing, ML tended to be better than MI. However, MI tended to be better in an unbalanced context, if the proportion of missing data was large. On the other hand, CC and IPW presented consistency in results. In general they had worse MSE, so they were not as efficient, particularly when the sample size is small. Thus, as a practical recommendation, we suggest to use MI in complex MAR scenarios in which the amount of missing data is large and the missing probability depends on the covariates. We suggest to use ML in scenarios with low missing proportion.

The results provided by the several methods in the case study agreed, both for the size of the estimated effects and for their standard errors. Variables IOTF and ABD in year 2000 were found significant for IOTF in 2006, while gender had no significance.

\section*{Disclosure statement}

The authors report there are no competing interests to declare.

\section*{Supporting}
Work supported by the Grant PID2020-118101GB-I00, Ministerio de Ciencia e Innovación (MCIN/ AEI /10.13039/501100011033).

\bibliographystyle{vancouver}
\bibliography{references}

\end{document}